\documentclass[%
 reprint,
nofootinbib,
 amsmath,amssymb,
 prb,
 showpacs,
 superscriptaddress
]{revtex4-2}

\usepackage{graphicx}
\usepackage{dcolumn}
\usepackage{bm}
\usepackage[normalem]{ulem}
\usepackage{tabularx}
\usepackage{comment}
\usepackage[colorlinks,citecolor=blue,linkcolor=blue,urlcolor=blue]{hyperref}
\usepackage[columnwise]{lineno}

\begin{document}


\title[Sample title]{Domain wall statics and dynamics in nanowires with arbitrary Dzyaloshinskii-Moriya tensors}

\author{Adriano Di Pietro}
\affiliation{ 
Istituto Nazionale di Ricerca Metrologica, Strada delle Cacce 91, 10135, Torino, Italy 
}%
 \affiliation{Politecnico di Torino, Corso Duca degli Abruzzi 24, 10129, Torino,  Italy.}

\author{Felipe García Sánchez}%
\affiliation{Departamento de Fisica Aplicada, Universidad de Salamanca, Salamanca 37008, Spain
}%

\author{Gianfranco Durin}%
\affiliation{Istituto Nazionale di Ricerca Metrologica, Strada delle Cacce 91, 10135, Torino, Italy
}%

\date{\today}

\begin{abstract}
The influence of different Dzyaloshinskii-Moriya interaction (DMI) tensor components on the static and dynamic properties of domain walls (DWs) in magnetic nanowires is investigated using one dimensional collective coordinates models and micromagnetic simulations. It is shown how the different contributions of the DMI can be compactly treated by separating the symmetric traceless, antisymmetric and diagonal components of the DMI tensor. First, we investigate the effect of all different DMI components on the static DW tilting in the presence and absence of in plane (IP) fields. We discuss the possibilities and limitations of this measurement approach for arbitrary DMI tensors. Secondly, the interplay of different DMI tensor components and their effect on the field driven dynamics of the DWs are studied and reveal a non-trivial effect of the Walker breakdown field of the material. It is shown how DMI tensors combining diagonal and off-diagonal elements can lead to a non-linear enhancement of the Walker field, in contrast with the linear enhancement obtainable in the usual cases (interface DMI or bulk DMI). 
\end{abstract}

\maketitle
\section{Introduction}

Recent years have seen an increased interest in the study of magnetic domain wall (DW) dynamics in perpendicularly magnetized nanowires as these are at the core of many emerging spintronic device concepts in memory storage \cite{Parkin2015,Kumar2022}, sensing \cite{Masciocchi2021,Klingbeil2021} and logic \cite{Luo2020,Allwood2005,Currivan2012}. To this day, many challenges still need to be addressed in order to make such technologies viable for the industry. Among the challenges to be faced is the phenomenon of Walker breakdown \cite{Glathe2008} field which sets a strong upper limit to the velocity a DW can be efficiently moved through a nanowire. It is a well established fact that magnetic DWs can be moved through a magnetic nanowire either via applied magnetic fields or spin transfer torque induced by spin polarized currents \cite{Thiaville2004,Bhowmik2015}. For small enough values of the driving force, the shape anisotropy of the material (which in thin film geometries favours Bloch walls) is able to counter act the torque on the magnetization that would cause precessional motion \cite{Mougin2007}. In the steady state regime, the DW is able to move rigidly and its peak velocity displays a linear dependence from the driving force. As the driving force increases, the competing torque becomes too strong and cannot be compensated by the effective field inside the DW: once that threshold, called Walker Breakdown (WB) field, is reached, the domain wall begins the so called precessional motion regime \cite{Thiaville2012,Mougin2007}, in which the peak velocity of the DW drastically reduces. Several strategies have been tried to counteract this phenomenon and increase the maximum attainable DW velocity \cite{Fattouhi2022,Lee2007}. For instance, the choice of materials displaying chiral interactions such as the Dzyaloshinskii-Moriya interaction (DMI) \cite{MOR-60,DZYALOSHINSKY1958241} in perpendicularly magnetized nanowires is known to greatly enhance the domain wall Walker breakdown \cite{Thiaville2012} because of the effective field component providing an additional restoring torque for the moving DW. While the effects of interface DMI (iDMI) are well known and understood, the effects of different, more exotic types of DMI \cite{Bogdanov1994,Leonov2016,Hoffmann2021} found in lower symmetry magnetic crystals are, to our knowledge, not studied in detail. The study of the possible effects induced by these additional DMI forms is becoming increasingly relevant as new deposition techniques are making the production of thin films with the required low symmetries a reality \cite{Karube2021,Swekis2021,Manna2018}. In the following we propose a micromagnetic study to analyze the DW statics and dynamics with additional terms accounting for arbitrary DMI tensors in magnetic nanowires. The paper is organized as follows: in Section \ref{Sec:2} we describe the energy contributions of our system and show how to compactly treat more exotic DMI tensors by decomposing them in antisymmetric, symmetric traceless and diagonal contributions. In Section \ref{Sec:3}, we introduce the collective coordinate models (CCMs) and derive the DW energy density for arbitrary DMI tensors both in the $q-\chi-\phi$ model \cite{Boulle2013} and the $q-\phi$ model \cite{Thiaville2002,Thiaville2012}. In Sections \ref{Sec:4} and \ref{Sec:5} we show how the derived energy densities correctly predict the DW tilting both with an without applied in-plane (IP) fields. In Section \ref{sec:6}, we explore the applicability of the canting angle method to measure forms of the DM tensor going beyond the iDMI discussed in ref.\cite{Boulle2013}. Finally, in Section \ref{Sec:7} we derive the dynamical equations for the DW in the $q-\phi$ model and show how the presence of certain combinations of DMI tensor components can lead to non-trivial changes in the DW Walker breakdown field. The derived analytical results are compared throughout with Micromagnetic simulations performed with the MuMax3 \cite{Vansteenkiste2014} software.  We conclude by summarizing our results and providing an outlook for future investigations in Section \ref{Sec:8}.

\section{Theoretical background}
\subsection{Energy density in the presence of arbitrary DMI tensors}\label{Sec:2}
\begin{figure*}[hbt]
    \centering
    \includegraphics[width=0.75\textwidth]{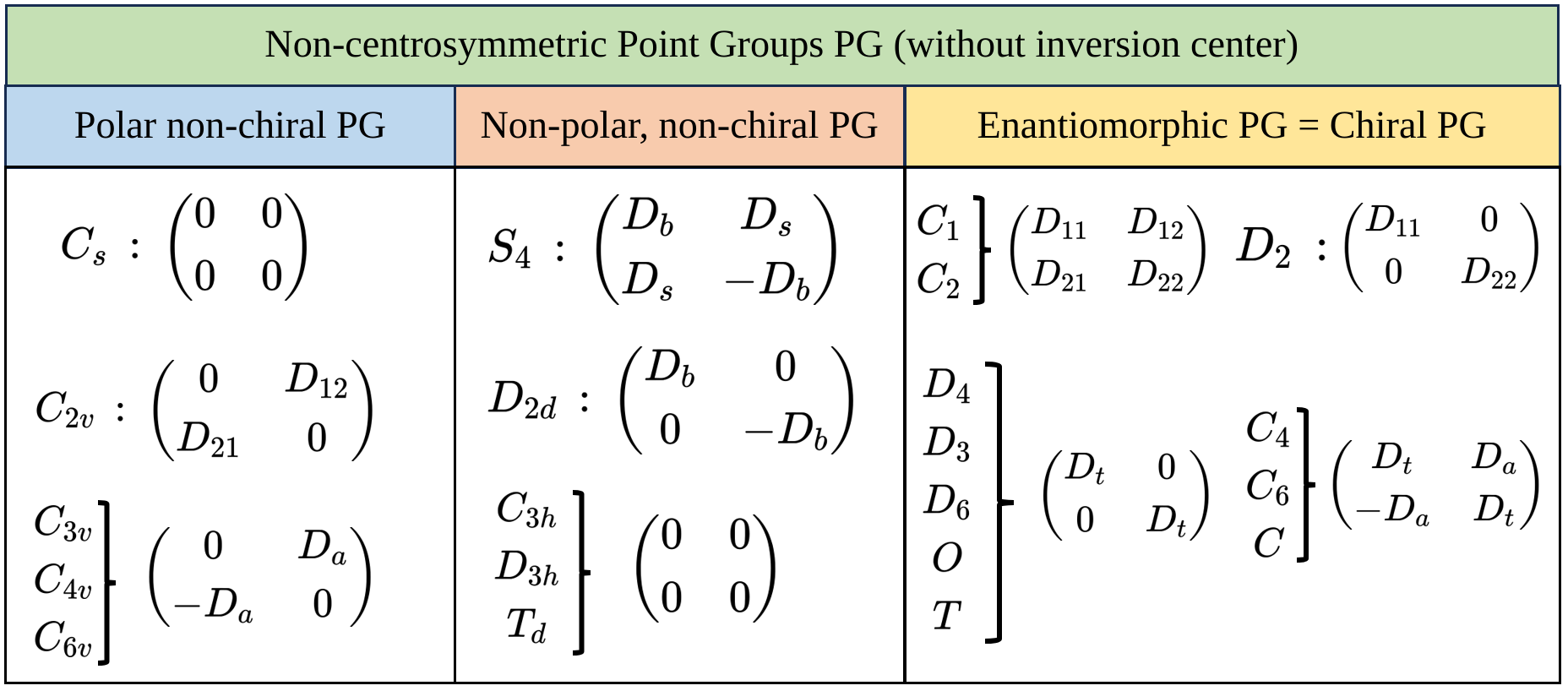}
    \caption{DMI tensor components for all 21 non-centrosymmetric crystallographic point
groups as imposed by the Neumann principle \cite{DiPietro2022}. The 11 centrosymmetric point groups have a vanishing DMI tensor
and are not shown. The components $D_a,D_s,D_b,D_t$ are the ones shown in the decomposition of eq.\eqref{eq:DMI_decomp}, while terms of the form $D_{ij}$ are combinations of $D_a,D_s,D_b,D_t$.}
    \label{fig:DMI_table}
\end{figure*}

We consider a magnetic ultrathin film of volume $\Omega_V$ grown on a substrate and a capping layer of a different material so that the symmetry is broken along the normal to the plane. In addition to the usual energy terms, we add a contribution relative to an arbitrary DMI tensor yielding a total density of the form \cite{Hoffmann,DiPietro2022}
\begin{align}
  E = \displaystyle \int_{\Omega_V}  \big\{ \, A |\bm{\nabla} \bm{m} |^2  -  Q_{ij}\mathcal{M}^{ji} - \frac{1}{2}\mu_0 M_s \bm{m}\cdot\bm{H}_d \nonumber \\ + K_u (1 - (\bm{m}\cdot\hat{\bm{u}}_z)^2) - \mu_0 M_s \bm{m}\cdot \bm{H}_z\, \big\} \, d^3 \bm{r}     \label{eq:main_res}
\end{align}
where $\bm{m}(x , t) = \bm{M}(x , t)/M_s$ is the normalized magnetization vector, $A$ is the symmetric exchange coefficient (in this case a constant), $\bm{H}_d$ is the magnetostatic field, $\bm{H}_z$ is the Zeeman field and $K_u$ is the uniaxial anisotropy constant with the easy axis directed along $z$. Finally, $Q_{ij}$ represents the DMI tensor and $\mathcal{M}_{ji} = \sum_k \varepsilon_{i k}(m_z \partial_j m_k - m_k \partial_j m_z  ) $ is the chirality of the magnetic configuration \cite{Hoffmann}. We remark how both the chirality $\mathcal{M}_{ji}$ and the DMI tensor $Q_{ij}$ treated here are already restricted to a 2 dimensional system, i.e. $\mathcal{M}_{ji} , Q_{ij} \in \mathbb{R}^{2 \times 2} $ and are reported in Fig.\ref{fig:DMI_table}. In the following we briefly outline some of the consequences of the symmetry properties of the DMI tensor. First of all, we remark that the DMI tensor, much like any other rank-2 tensor, can be decomposed in a sum of symmetric traceless, antisymmetric and diagonal components as follows
\begin{equation}
   \hat{\bm{Q}}_{ij} = \underbrace{\begin{pmatrix} 0 & D_a  \\ -D_a & 0\end{pmatrix}}_{\text{Antisymmetric}} + \underbrace{\begin{pmatrix} D_b & D_s  \\ D_s &  -D_b \end{pmatrix}}_{\text{Symmetric-traceless}} + \underbrace{\begin{pmatrix} D_{t} & 0  \\ 0 & D_{t} \end{pmatrix}}_{\text{Diagonal}}  \label{eq:DMI_decomp}.
\end{equation}  
A purely anti-symmetric DMI tensor $(Q_A)_{ij} = \sum_k D_k \varepsilon_{kij}$ yields Lifshitz invariant energy density terms of the form 
\begin{equation}
    \mathcal{E}_{A;DMI} = -2 \bm{D}\cdot [ \bm{m}(\nabla \cdot \bm{m}) - (\nabla \cdot \bm{m})\bm{m}] \label{iDMI},
\end{equation}
which correspond to the interface DMI (iDMI) term often studied in the literature \cite{Thiaville2012,Miron2011}. The symmetric component of the DMI tensor, on the other hand, yields an energy contribution of the form
\begin{equation}
    \mathcal{E}_{S;DMI} = - \bm{m}\cdot(\hat{\bm{Q}}_S\nabla \times \bm{m}),
\end{equation}
where $\hat{\bm{Q}}_S\nabla = \sum_j(Q_S)_{ij}\partial_j$. A DMI of this form is related to the so called "anisotropic DMI" in the discrete microscopic treatment \cite{Ga2022,Camosi2017,Cui2022}. The special case of a purely diagonal matrix yields an energy term of the form
\begin{equation}
    \mathcal{E}_{S;DMI} = - 2 (Q_S)_{ii}(\bm{m} \cdot \partial_i \bm{m})_i
\end{equation}
which, in the case of a single independent component $Q_{ii} = D$ yields 
\begin{equation}
    \mathcal{E}_{S;DMI} = - 2 D \, \bm{m} \cdot (\nabla \times \bm{m}). 
\end{equation}
This energy contribution corresponds to a bulk DMI (bDMI) term responsible for stabilizing bulk chiral structures \cite{Nagaosa2013}. 

\subsection{Collective coordinate models with arbitrary DMI}\label{Sec:3}
\begin{figure*}[hbt]
    \noindent
    \includegraphics[width=\textwidth]{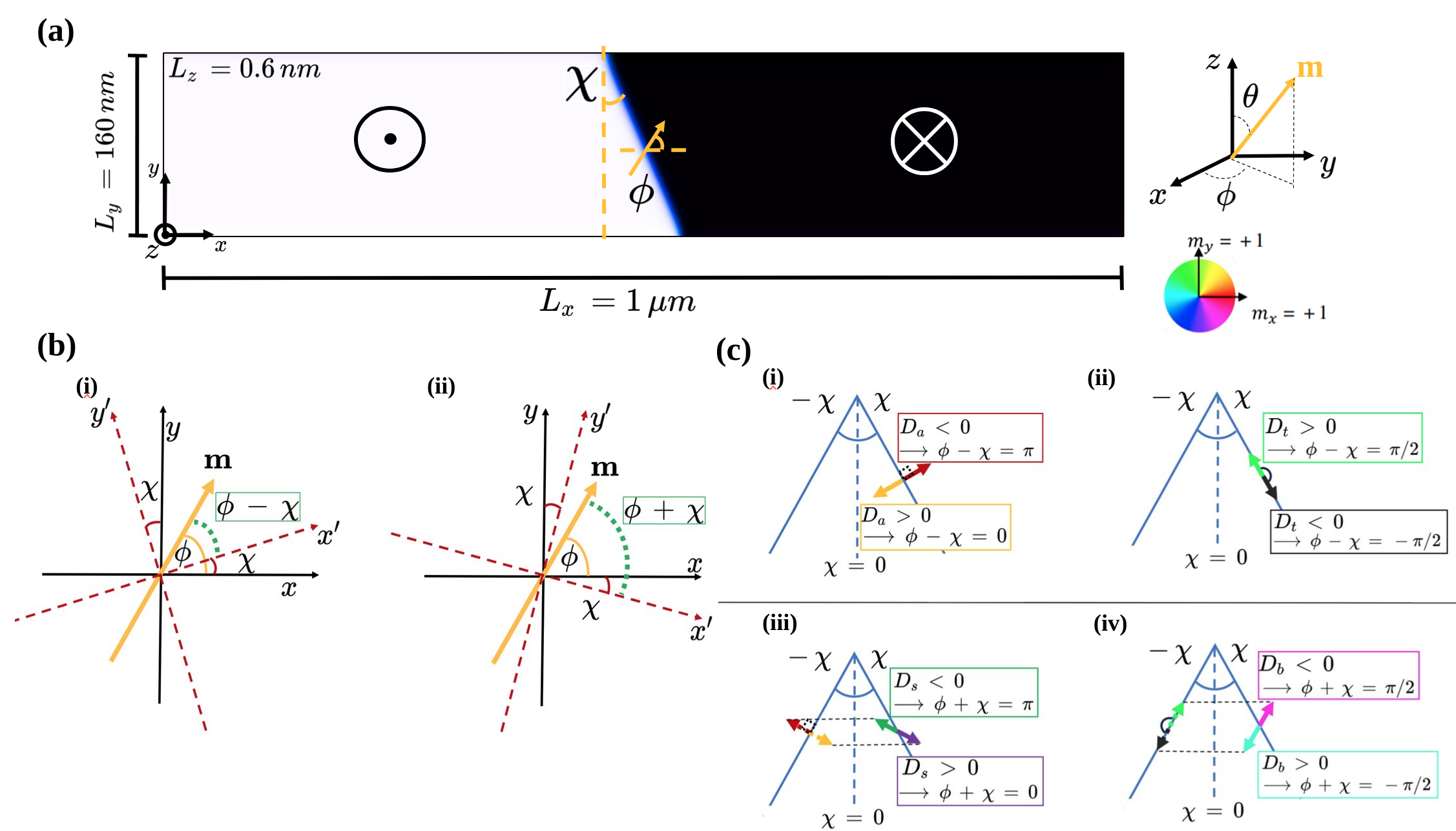}
    \caption{(a) Scheme of the system used in the micromagnetic simulations. We show the dimensions $L_x = 1 ~ \mu$m , $L_y = 160$ nm and $L_z = 0.6$ nm as well as the internal DW angle $\phi$ and the DW tilt angle $\chi$. (b) $\phi$ and $\chi$ angles in the case of $\chi >0$ (i.) $\phi$ and $\chi$ angles in the case of $\chi >0$ (ii.) (c)  Schematic representation of the internal DW angle stabilized by the presence of the different DMI tensor components of eq.\eqref{eq:E_dens_tilt} in the presence of an applied IP field}
    \label{fig:Setup}
\end{figure*}

Since the contributions of the iDMI terms (i.e. the $D_a$ part of eq.\eqref{eq:DMI_decomp}) and bDMI (i.e. the $D_t$ part of eq.\eqref{eq:DMI_decomp}) to ordinary collective coordinate models (CCMs) are known \cite{Boulle2013,Li2022}, to account for the complete DMI tensor we just have to compute the energy density terms relative to the symmetric $D_s$ and traceless $D_b$ components. To this end, we consider a DMI tensor compatible with the $S_4$ point group symmetry which has the form 
\begin{equation}
    \hat{\bm{Q}}_{S_4} = \begin{pmatrix} D_b & D_s  \\ D_s &  -D_b \end{pmatrix}.
\end{equation}
Plugging this DMI tensor in eq.\eqref{eq:main_res} and writing the magnetization in spherical coordinates $\bm{m} = \bm{M}/M_s = \big(\sin{\theta}\cos{\varphi} , \sin{\theta}\sin{\varphi} , \cos{\theta}\big)^T$  we can write the $S_4$ DMI energy density as follows
\begin{align}
    \mathcal E_{DMI, S_4} = D_b(\sin \varphi ~ \partial_x \, \theta  + \cos \varphi ~ \partial_y \theta) + \nonumber \\ D_s (\sin \varphi ~ \partial_y \theta - \cos \varphi ~ \partial_x \theta ). \label{eq:e_dens_S4}
\end{align}
To derive the CCM we must now substitute $\theta$ and $\phi$ with the Ansatz for the tilted DW \cite{Boulle2013}
\begin{align}
    \tan\left(\frac{\theta(q , \chi)}{2}\right) &= \exp\left( Q \frac{(x-q)\cos\chi + y \sin \chi}{\Delta}\right) \label{eq:theta_tilt} \\ \varphi(t) &= \phi(t) , 
\end{align}
where $q$ represents the DW position along the $x$-axis, $\chi$ represents the DW tilting angle, $\Delta$ the DW width and $Q = \pm 1$ represents the sense of rotation of angle $\theta$ ( i.e. $Q = \pm 1 \Rightarrow m_z(- \infty) = \pm 1 ~ $ and $m_z(+ \infty) = \mp 1 $ ). For a schematic of the system and the angles, refer to Fig.\ref{fig:Setup}-(a). Noticing that the Ansatz of eq.\eqref{eq:theta_tilt} allows us to compactly compute the derivatives of eq.\eqref{eq:e_dens_S4} as 
\begin{align}
    \partial_x \theta = Q \frac{\sin \theta \cos \chi}{\Delta} \\ \partial_y \theta = Q \frac{\sin \theta \sin \chi}{\Delta}  
\end{align}
we can write the energy density $\mathcal{E}_{DMI, S_4}$ of eq.\eqref{eq:e_dens_S4} as 
\begin{align}
\mathcal E_{DMI, S_4} = Q \frac{\sin \theta}{\Delta} \big[D_b \sin(\phi + \chi) - D_s \cos (\phi + \chi)\big]. \label{eq:e_dens_S4_coord}  
\end{align}
To obtain a DW surface energy, we quench the $x$-degree of freedom of eq.\eqref{eq:e_dens_S4_coord} by integrating it out
\begin{align}
     \sigma_{DW, S_4} &= \displaystyle{\int_{-\infty}^{+ \infty}} \mathcal E_{DMI, S_4} ~ \text{d}x \nonumber \\ &=   \pi Q \big[ D_b \sin(\phi + \chi)- D_s \cos (\phi + \chi)\big] \label{eq:sigma_dens_S4_tilt}.
\end{align}
We can now add this DW energy component to the other energy terms already used in \cite{Boulle2013} to obtain a generalized DW energy density as a function of all the DMI tensor components
\begin{widetext}
\begin{align}
     \sigma_{DW}(\phi,\chi) = &2 \frac{A}{\Delta} + \pi Q \big[D_a \cos(\phi - \chi) - D_s \cos (\phi + \chi) - D_t \sin(\phi - \chi) + D_b \sin(\phi + \chi)\big] + \nonumber \\ &2 \Delta (K_0 + K \sin^2(\phi - \chi)) - \pi \Delta M_s (H_y \sin \phi + H_x \cos \phi ) \label{eq:E_dens_tilt}, 
\end{align}
\end{widetext}

with $K_0 = K_u + \frac{M_s \mu_0}{2}(N_x - N_z)$ and $K = \frac{M_s \mu_0}{2}(N_y - N_x) $ being the effective and shape anisotropy constants, respectively. $N_x , N_y , N_z$ are the demagnetizing factors which depend on the geometry of the sample \cite{Coey2001,Aharoni1998}. If the phenomenon of DW tilting is not to be considered, the properties of the DW can be studied by considering the more simple $q-\phi$ model \cite{Thiaville2002,Thiaville2012} which can be obtained by setting $\chi = H_x = H_y = 0$ in eq.\eqref{eq:E_dens_tilt}, 
\begin{align}
     \sigma_{DW}(\phi) = &2 \frac{A}{\Delta} + \pi Q \big[(D_a - D_s) \cos (\phi) + (D_b - D_t) \sin(\phi)\big] + \nonumber \\ &2 \Delta (K_0 + K \sin^2(\phi))
     \label{eq:E_dens_non_tilt_no_fields}
\end{align}

\section{Results}

\subsection{In plane field driven DW titling in the presence of arbitrary DMI tensors}\label{Sec:4}
\begin{figure}[hbt]
    \noindent
    \includegraphics[width=0.5\textwidth]{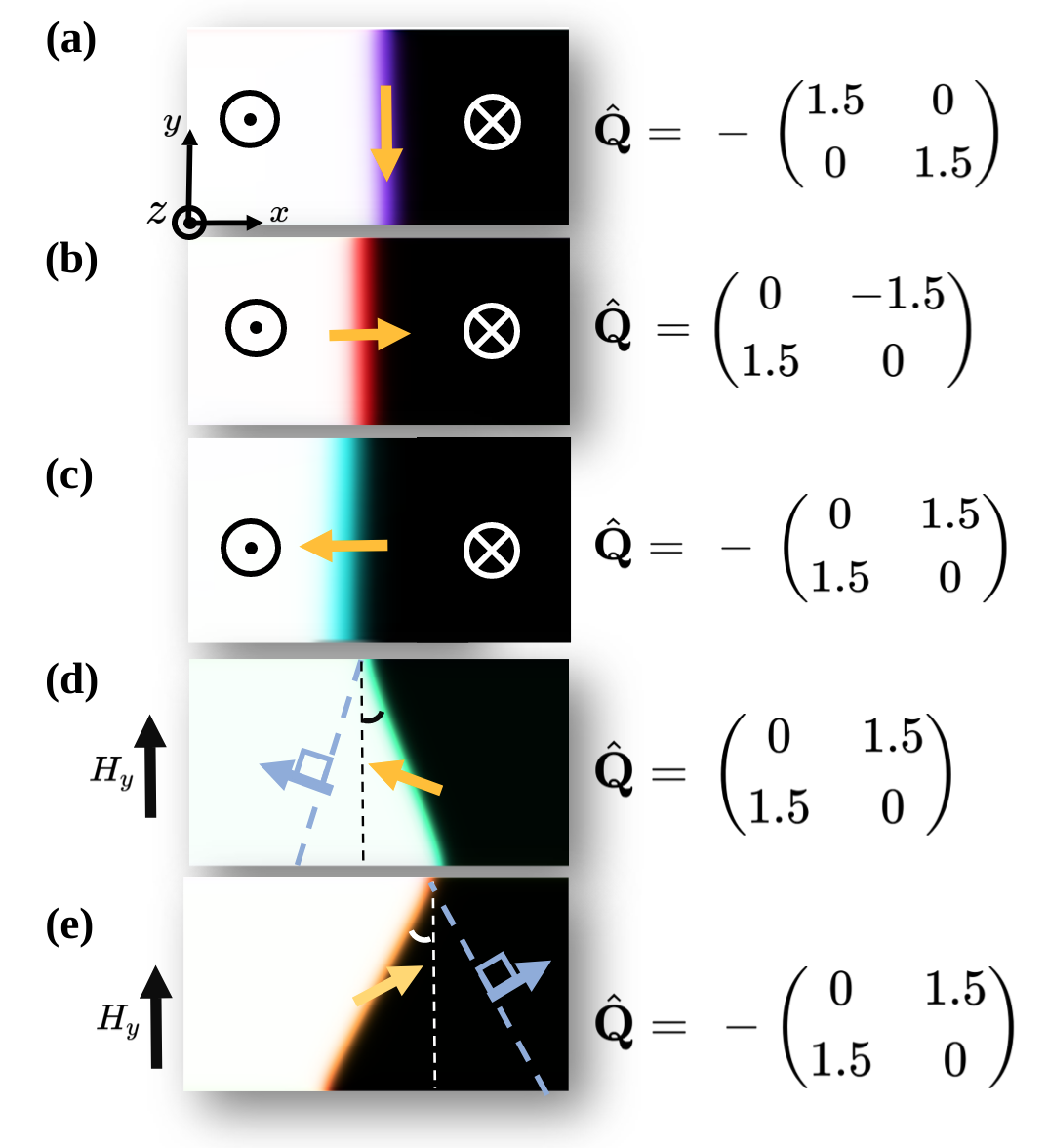}
    \caption{a-c) Internal DW magnetization angle stabilized by 3 different representative DMI tensors in the absence of an applied IP field (d-e) Internal DW magnetization angle stabilized by 2 different representative DMI tensors in the presence of an applied IP field.  The mirrored image of the tilted domain wall in (d) and (e) is included for clarity. The DMI tensor components are expressed in mJ/m$^2$.}
    \label{fig:Components_analysis}
\end{figure}

\begin{figure*}[hbt]
    \noindent
    \includegraphics[width=\textwidth]{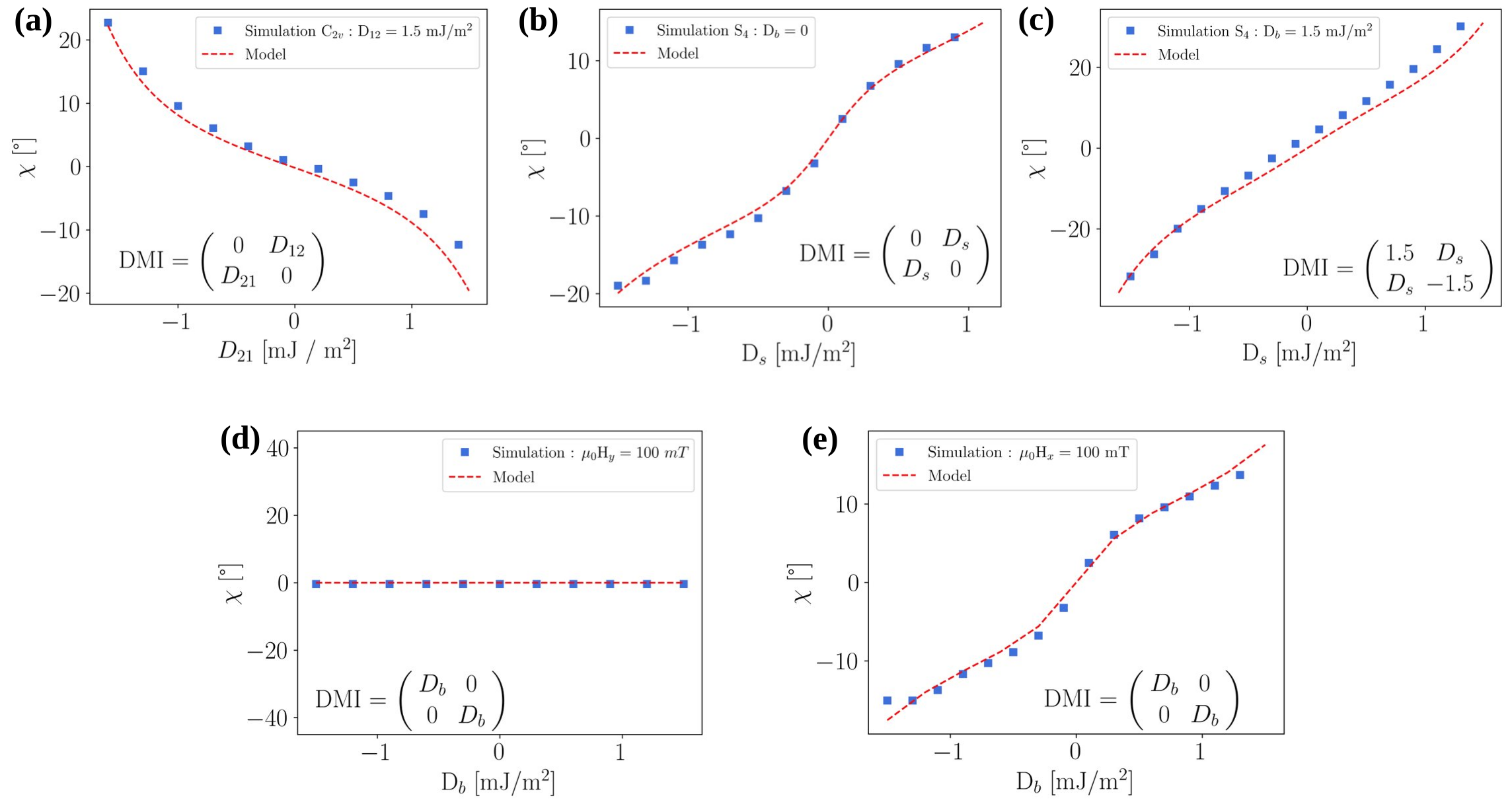}
    \caption{ Comparison of Micromagnetic simulations \cite{Vansteenkiste2014} and numerical minimization of the energy density of eq.\eqref{eq:E_dens_tilt} for (a) The tilting angle $\chi$ as a function of the D$_{21}$ DMI tensor component of in the case of a C$_{2v}$ symmetric DMI. (b) Tilting angle $\chi$ as a function of the D$_{s}$ DMI tensor component of in the case of a S$_4$ symmetric DMI in the case of D$_b = 0$. (c) Tilting angle $\chi$ as a function of the D$_{s}$ DMI tensor component of in the case of a S$_4$ symmetric DMI in the case of D$_b = 1.5$ mJ/m$^2$. (d) Tilting angle $\chi$ as a function of the D$_{b}$ DMI tensor component of in the case of a T symmetric DMI with an IP applied field in the y-direction of magnitude $\mu_0 H_y = 100 $ mT (e) Tilting angle $\chi$ as a function of the D$_{b}$ DMI tensor component of in the case of a T symmetric DMI with an IP applied field in the x-direction of magnitude $\mu_0 H_x = 100$ mT. }
    \label{fig:simulations}
\end{figure*}

It is a well established fact, that the presence of iDMI induces a tilt in the DW profile \cite{Boulle2013,Nasseri2018} under the application of an external in-plane (IP) transverse field. The origin of this phenomenon is explained by considering the relative energy balance of the DW in presence of chiral interactions and Zeeman fields. In the absence of applied IP fields, the DW reaches an internal equilibrium angle dictated by the relative strength of DMI and demagnetizing contributions. If we apply an external IP field along the positive $y$-direction, the DW magnetization is going to feel the added competing interaction requiring it to align along the direction of the external field. At the same time, the iDMI produces an effective field component that stabilizes Néel walls. To try and accommodate both torques, the DW tilts by an angle $\chi$ increasing the DW energy by a factor $1 / \cos \chi$. In the following we try and extend what is known about DW tilting in the presence of iDMI to the case of arbitrary DMI tensors (see Fig.\ref{fig:DMI_table}).  As a first step, we analyze the new DMI energy terms of the $\chi = 0$ case 
\begin{equation}
    \sigma_{DMI} = \pi\big[(D_a-D_s) \cos(\phi) + (D_b - D_t) \sin(\phi) \big]. \label{eq:sigma_DM_non_tilt}
\end{equation}
and of the $\chi \neq 0$ case
\begin{align}
    \sigma_{DMI} = \pi\big[D_a \cos(\phi - \chi) - D_s \cos (\phi + \chi) \nonumber \\ - D_t \sin(\phi - \chi) + D_b \sin(\phi + \chi)\big], \label{eq:sigma_DM_tilt}
\end{align}  
where we have set $Q = 1$ for convenience. In the untilted case $\chi = 0$, eq.\eqref{eq:sigma_DM_non_tilt} suggests that the different DMI tensor components all simply induce either Néel or Bloch wall stabilizing effective fields, however this intuitive picture is only valid as long as no tilting is observable. If tilting is present (eq.\eqref{eq:sigma_DM_tilt}) in the system, we need to take in account the fact that the $D_s$ and $D_b$ components minimize the energy of the DW as a function of $\phi + \chi$ as opposed to $\phi - \chi$. As a first step to understand the implications of this difference, we discuss the equilibrium angles stabilized by all the different DMI tensor components of eq.\eqref{eq:E_dens_tilt}.  The values of the physical parameters used in the micromagnetic simulation for the statics and dynamics of the DW represent the values measured in Pt/Co/AlOx nanowires \cite{Miron2011}. We set the exchange constant $A = 10^{-11} $J/m, the saturation magnetization $M_s = 1.09 $ MA/m, the effective anisotropy constant $K_0 = 1.25$ MJ/m$^3$, the damping coefficient $\alpha = 0.5$. The  chosen nanowire dimensions are $L_x = 1 ~ \mu$m , $L_y =  160 $ nm and $L_z = 0.6$ nm (see Fig.\ref{fig:Setup}-(a) for the schematics of the setup).
By observing the Fig.\ref{fig:Setup}-(b)-i., (assuming $\chi > 0$) we notice how $\phi - \chi$ represents the DW magnetization angle in the reference frame of the tilted DW. $\phi + \chi$ on the other hand, represents the DW magnetization in the reference frame of a mirrored image of the tilted DW, i.e. with a canting angle of $-\chi$ (see Fig.\ref{fig:Setup}-(b)-ii). As obtained from \cite{Li2022} and \cite{Boulle2013}, the $D_a$ and $D_t$ components of the DM tensor stabilize, respectively, Néel and Bloch DWs in the reference frame of the tilted DW (see Fig.\ref{fig:Setup}-(c)-i. and -ii.). On the other hand, the dependence from the $\phi + \chi$ angle of $D_s$ and $D_b$ components results in the stabilization of Néel or Bloch DWs in a reference located in a mirror image version of the DW itself (see Fig.\ref{fig:Setup}-(c)-iii. and -iv.). 
To emphasize how the effect of the $D_s$ and $D_b$ components can only be distinguished from the $D_a$ and $D_t$ contributions in the presence of DW tilting (i.e. $\chi \neq 0$), we analyze the equilibrium configurations obtained from the minimization of the untilted case and compare them with micromagnetic simulations performed with a version of the MuMax3 code \cite{Vansteenkiste2014} suitably modified to account for the new components of the DMI tensor of eq.\eqref{eq:e_dens_S4_coord}. By observing eq.\eqref{eq:E_dens_non_tilt_no_fields}, it is immediately apparent that in the case $\chi = 0$, the effect of $D_s$ and $D_a$ (or $D_b$ and $D_t$) cannot be untangled as all these energy terms contribute to the stabilization of an untilted Néel- ($D_s$ and $D_a$) or an untilted Bloch-wall ($D_b$ and $D_t$). This effect is clearly visible in Fig.\ref{fig:Components_analysis}-(a), where a DM tensor composed only of a $D_a$ part stabilizes a Néel wall (Fig.\ref{fig:Components_analysis}-(a)-ii.) while a DM tensor composed of a $D_s$ part stabilized a Néel wall with opposite chirality (Fig.\ref{fig:Components_analysis}-(a)-i.). In the presence of DW tilting (induced e.g. by the presence of an applied IP field along the $y$-axis), the different energy contributions become distinguishable as can be seen in Fig.\ref{fig:Components_analysis}-(b)-i. and -ii., where the DW magnetization in the presence of $D_s = \pm 1.5$ mJ/m$^2 ,~ D_a = D_b = 0$ and an IP field of $H_y = 100$ mT points in a direction compatible with a Néel wall in a reference frame tilted in the opposite direction $- \chi$. 
(see the dotted line in Fig. 3(d-e)
The simultaneous presence of all the different DMI contributions as well as their relative importance is more complex and is studied both numerically, via the minimization of eq.\eqref{eq:E_dens_tilt} and with micromagnetic simulations.
In Fig.\ref{fig:simulations}-(a) we observe the tilting angle $\chi$  of the DW in the presence of a DMI tensor compatible with $C_{2v}$ crystal symmetry \cite{Bogdanov1994}, i.e.
\begin{equation}
    \hat{\bm Q}_{C_{2v}} = \begin{pmatrix} 0 & D_{12}  \\ D_{21} & 0 \end{pmatrix} . \label{eq:C_2v}
\end{equation}
By observing the value of $\chi$ for $D_{21} = 0$ we notice a vanishing of the DW tilting while a form of DMI ($D_{12} \neq 0$) is still present. This phenomenon can be understood using the intuitive picture of competing effective fields. As can be observed in the untilted model of eq.\eqref{eq:E_dens_non_tilt_no_fields}, the term stabilizing Nèel walls has the form $(D_a - D_s) \cos(\phi)$. In the $C_{2v}$ case of eq.\eqref{eq:C_2v}, we have $D_a = (D_{12} - D_{21})/2$ and $D_s = (D_{12} + D_{21})/2 $  and therefore
\begin{equation}
    \Rightarrow D_a - D_s = D_{21}
\end{equation}
implying that the component of the DMI tensor that stabilizes Néel walls (and is responsible for tilting since it competes with the $H_y$ torque) is the $D_{21}$ component.  In Fig.\ref{fig:simulations}-(b,c) on the other hand, we observe the behavior or the DW tilting angle $\chi$ in the presence of a DMI tensor compatible with $S_4$ crystal symmetry \cite{Bogdanov1994,Leonov2016} in 2 different cases. In Fig.\ref{fig:simulations}-(b) we have
\begin{equation}
 \hat{\bm{Q}}_{S_4} =  \begin{pmatrix} 0 & D_{s}  \\ D_{s} & 0 \end{pmatrix} ,
\end{equation}
while in Fig.\ref{fig:simulations}-(c) we have
\begin{equation}
 \hat{\bm{Q}}_{S_4} =  \begin{pmatrix} D_{b} & D_{s}  \\ D_{s} & -D_{b} \end{pmatrix} .\label{eq:DMI-S_4}
\end{equation}
By comparing the 2 graphs we can observe how the presence of $D_b$ terms emphasizes the canting effect. This can be understood by recalling how the $D_b$ terms energetically favors the formation of Bloch walls. In the presence of a transverse field along the $y$-direction, the effective field coming from $D_b$ acts constructively and exacerbates the canting one would normally observe without $D_b$. In Fig.\ref{fig:simulations}-(d,e) we explore the canting angle $\chi$ in the presence of a DMI tensor compatible with the point group symmetry $T$ (or others \cite{DiPietro2022}) i.e. 
\begin{equation}
\hat{\bm{Q}}_{T} =  \begin{pmatrix} D_{t} &  0  \\ 0 & D_{t} \end{pmatrix}.    
\end{equation}
In Fig.\ref{fig:simulations}-(d,e) we study the behavior of $\chi$ as a function of $D_t$ in the presence of a transverse field along the $y$-direction (Fig.\ref{fig:simulations}-(d)) and in the presence of a transverse field along the $x$-direction (Fig.\ref{fig:simulations}-(e)). We observe how tilting is only present in the case of an applied transverse field applied along the $x$-direction. This can be explained observing eq.\eqref{eq:E_dens_tilt} where we notice that the DMI associated to $D_t$ tends to stabilize Bloch walls (Fig.\ref{fig:Components_analysis}-(c)-ii.): as a consequence a transverse $H_x$ field tries to change the internal DW magnetization to a Néel configuration. Much like in the case of the $D_a$ and $D_s$ (see Fig.\ref{fig:Setup}-(c)-i. and -iii.), the DW responds by tilting to try and accommodate both the Zeeman- and the $D_t$ effective field.

\subsection{Intrinsic DW tilting in the presence of $D_b$ and $D_s$}\label{Sec:5}
\begin{figure*}[hbt]
    \noindent
    \includegraphics[width=0.9\textwidth]{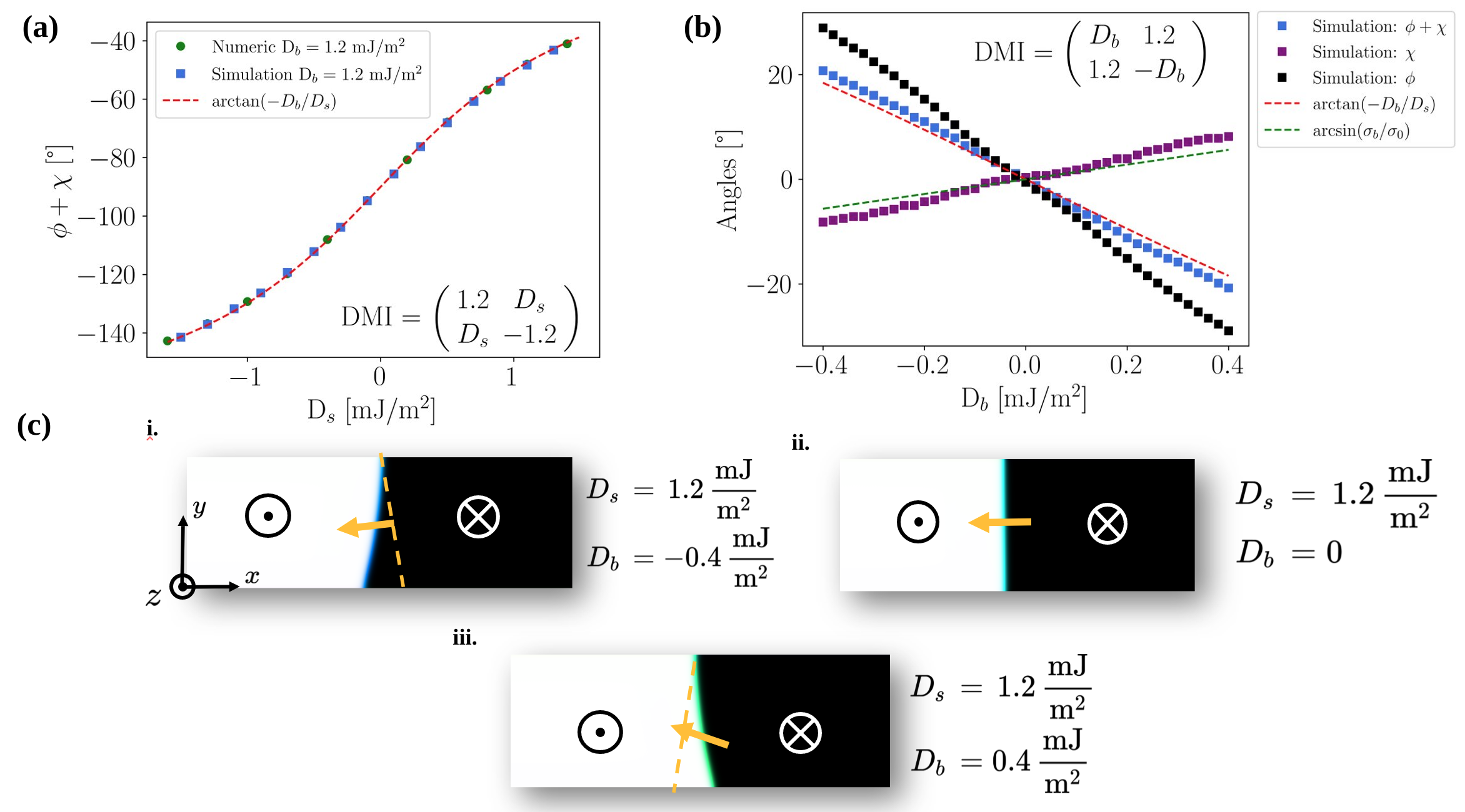}
    \caption{(a) DW angle $\phi + \chi$ as a function of the off-diagonal DMI tensor components $D_s$ in the $S_4$ symmetric case.  (b) DW angles as a function of the diagonal DMI tensor components $D_b$ ( $D_b/D_s \ll 1$ limit) in the $S_4$ symmetric case. (c) intrinsic DW tilting in the presence of simultaneous presence of $D_s$ and $D_b$ for $3$ representative cases.}
    \label{fig:Intrinsic_DMI} 
\end{figure*}

As mentioned in the discussion of Sec.\ref{Sec:4}, the appearance of DW tilting in perpendicularly magnetized nanowires is a consequence of the internal equilibrium of torques trying to orient the DW magnetization some preferred configuration. According to eq.\eqref{eq:E_dens_tilt}, if the DMI tensor of the system displays both diagonal and off diagonal components, the conflict of Néel- and Bloch-wall stabilizing torques is expected to be present even in the absence of an applied IP field. By observing Fig.\ref{fig:Intrinsic_DMI}-(c), we can in fact see how the presence of a DMI tensor compatible with the $S_4$ point group symmetry (see eq.\eqref{eq:DMI-S_4}), DW tilting occurs even in the absence of IP fields. In the thin film limit, considering a situation where the DMI strength dominates the demagnetizing field, the magnetization angle in the reference frame of the DW (i.e. $\phi + \chi$) can be easily derived by minimizing the 
simplified DW energy density
\begin{equation}
\sigma_{DW}(\phi,\chi) = 2 \frac{A}{\Delta} + \pi\big[ D_b \sin(\phi + \chi) - D_s \cos (\phi + \chi)\big],
\end{equation}
which yields the simple solution (Fig.\ref{fig:Intrinsic_DMI}-(a))
\begin{equation}
        \phi + \chi = \arctan \left(- \frac{D_b}{D_s} \right). 
\end{equation}
To obtain an approximate solution for the tilting angle $\chi$ in the $D_s / D_b \ll 1$ limit as a function of the material parameters, we can follow the procedure outlined in ref.\cite{Boulle2013} making the analogy between the $D_b$ DMI field  and an applied field along the $y-$axis. As discussed in Section.\ref{Sec:2}, DW tilting is the result of an energy balance between satisfying the internal constraints of the DW and the energy cost due to its surface area increase. We imagine a scenario where the initial state of the DW is a Néel configuration (large $D_s$ hypothesis), i.e. $\sigma_0 = 2A / \Delta + \pi D_s + 2 \Delta K_0$. The energy of the DW surface scales with $\sim 1 / \cos \chi$, while the energy gain of the $D_b$ DMI component in the DW scales approximately with $\sin \chi$. If we assume a small $D_b$ contribution ($D_b / D_s \ll 1$), we can approximate the DW energy in the Néel configuration as fixed and the energy of the DW as 
\begin{equation}
    \sigma_{DW} \approx \frac{\sigma_0 - \pi D_b \sin{\chi} }{\cos \chi},
\end{equation}
which is minimized by
\begin{equation}
    \sin \chi = \frac{\pi D_b}{\sigma_0} =  \frac{\pi D_b}{2A / \Delta + \pi D_s + 2 \Delta K_0}.
\end{equation}
As we can see from Fig.\ref{fig:Intrinsic_DMI}-(b), the above formula fits the simulations data reasonably well for small $D_b$, where the dependency of the tilting angle $\chi$ from the $D_b$ component is approximately linear.

\subsection{Measuring $D_a$ and $D_s$ DMI contributions with IP fields} \label{sec:6}

\begin{figure}[hbt]
    \noindent
    \includegraphics[width=0.5\textwidth]{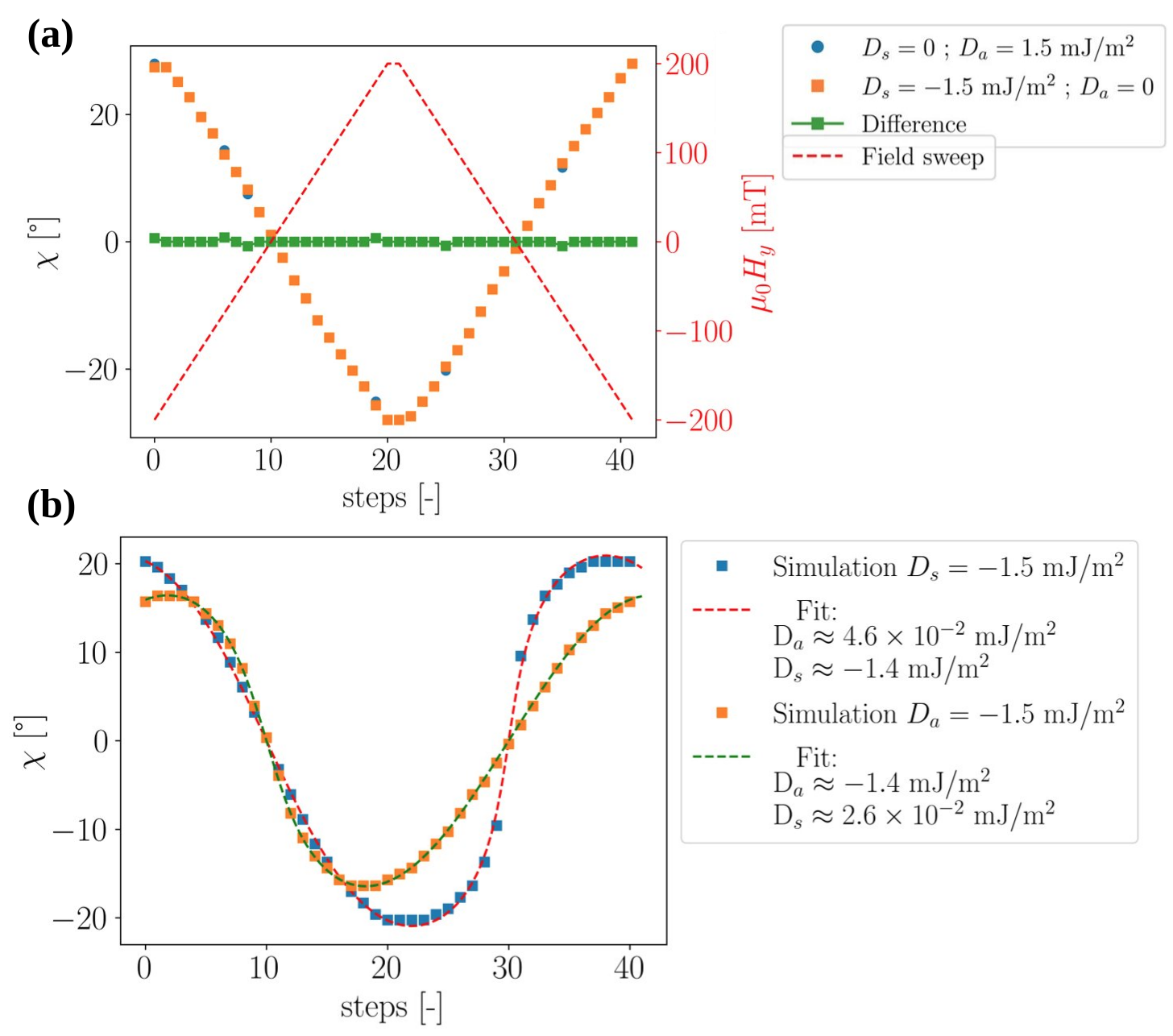}
    \caption{(a) DW tilting angle $\chi$ response to an H$_y$ field sweep from $-200$ mT to $+200$ mT in the case of pure $D_a$ (blue dots) and pure $D_s$ (orange squares) contributions to DMI. As can be seen the $2$ responses overlap almost completely. (b) DW tilting angle $\chi$ response to a rotating IP field with H$_y$ and H$_x$ components (see eq.\eqref{eq:rot_field} in the case of pure $D_a$ and pure $D_s$ contributions to DMI. The dashed curves are obtained by fitting the energy minimum of eq.\eqref{eq:E_dens_tilt} onto the results obtained via micromagnetic simulations using $D_a$ and $D_s$ as the fitting parameters.}
    \label{fig:Da_Ds_measures}
\end{figure}

According to the discussion of Sec.\ref{Sec:4} and Fig.\ref{fig:Components_analysis}, it might seem impossible to use the canting angle as a function of applied IP fields to measure $D_a$ and $D_s$ since in the untilted case of eq.\eqref{eq:E_dens_non_tilt_no_fields}, the $D_s$ energy density component simply contributes to the stabilization of a Néel wall and can either collaborate or compete with the $D_a$ contribution depending on the relative sign. We can in fact observe how in Fig.\ref{fig:Da_Ds_measures}-(a), the response of the tilting angle $\chi$ to an IP H$_y$ field in the case of $D_s \neq 0$ is identical to the case $-D_a$ and cannot be distinguished. However, according to eq.\eqref{eq:E_dens_tilt}, Fig.\ref{fig:Components_analysis}-(b)-i. and -ii., even when the canting angle $\chi$ is identical, the equilibrium angle $\phi$ inside the DW in the presence of $D_s$ is different when compared to a system with $D_a$. This implies that the simultaneous action of H$_y$ and H$_x$ IP fields should induce a different response of the DW canting angle $\chi$ in the nanowire. In Fig.\ref{fig:Da_Ds_measures}-(b), we show how the canting angle $\chi$ responds differently in the presence of $D_a$ or $D_s$ under the application of a rotating IP field of the form
\begin{equation}
    \mu_0\bm{H}  = \mu_0 H_0\begin{pmatrix}
        \cos(\omega t) \\ \sin(\omega t) \\ 0
    \end{pmatrix}, \label{eq:rot_field}
\end{equation}
where $t \in [ 0 , T] ~ , ~ \omega = 2\pi/T$ and $\mu_0 H_0 = 100 mT$. We stress the fact that the variable $t$ does not have the unit of a physical time, since in the simulation the canting angle $\chi$ in response to the applied field is recorded after the system has had time to relax and not after a fixed time interval. In the x-axis of Figs-\ref{fig:Da_Ds_measures}-(a,b) we refer to this variable as "steps". In Fig.\ref{fig:Da_Ds_measures}-(b) also shows how the form of these curves could in principle be fitted to Eq.\eqref{eq:E_dens_tilt} to extract the $D_a,D_s$ coefficients, potentially allowing for the magneto-optical measurements of different DMI tensor components with the canting angle method. The fit is performed by calculating $\chi$ from a constrained minimization of the DW energy density of eq.\eqref{eq:E_dens_tilt} using $H_x,H_y$ as variables and $D_a$ and $D_s$ as fitting parameters.

\subsection{Domain-wall dynamics in the presence of arbitrary DMI tensors}\label{Sec:7}
After having studied the effects of the different components of the DMI tensor on the static configurations of magnetic domain walls in nanowires, we now focus on the effects on the dynamics.
Given  that in the field driven, steady state regime the magnetization angle in the reference frame of the DW is only dependent on the IP torques exerted by the driving field $H_z$, the anisotropy contributions $H_k$ and the various components of the DMI tensor, we can avoid considering $\chi$ as a collective coordinate in the dynamical equations and work with the simpler $q-\phi$ model \cite{Boulle2013,Nasseri2018} whose DW energy density $\sigma_{DW}(q,\phi) $ with the generalized chiral interaction tensor from eq.\eqref{eq:DMI_decomp} can be written as 
\begin{align}
     \sigma_{DW}(\phi) = &2 \frac{A}{\Delta} + \pi\big[(D_a - D_s) \cos (\phi) + (D_b - D_t) \sin(\phi)\big] + \nonumber \\ &2 \Delta (K_0 + K \sin^2(\phi)) - \pi \Delta M_s (H_y \sin \phi + H_x \sin \phi ).
     \label{eq:E_dens_non_tilt}
\end{align}
By explicitly writing the Lagrangian of the DW as $\mathcal{L} = \sigma_{DW} + (M_s / \gamma)\phi \dot{\theta}\sin{\theta}$ and the Rayleigh dissipation function to correctly account for damping effects $\mathcal{F} = (\alpha M_s /2\gamma) ~ \dot{\bm{m}} $. We can derive the equations of motion from the Euler-Lagrange-Rayleigh equation \cite{Thiaville2002}, 
\begin{align}
    \frac{\partial \mathcal{L}}{\partial X}-\frac{d}{d t}\left(\frac{\partial \mathcal{L}}{\partial \dot{X}}\right)+\frac{\partial \mathcal{F}}{\partial \dot{X}}=0 ,~~ X \in \{ q , \phi , \Delta \},
\end{align}
obtaining the following equations of motion

\begin{align}
 \dot{q} =&\frac{\Delta \gamma_0}{1+\alpha^2}\bigg[\alpha Q H_z+Q H_K \frac{\sin 2 \varphi}{2}-\frac{\pi}{2} \tilde{f}^\prime_{\mathrm{DMI}}(\varphi) \nonumber \\  & -Q \frac{\pi}{2}\left(H_y \cos \varphi-H_x \sin \varphi\right)\bigg], \\  \dot{\varphi} =&\frac{\gamma_0}{1+\alpha^2}\bigg[H_z-\alpha\bigg(H_K \frac{\sin 2 \varphi}{2}- Q \frac{\pi}{2} \tilde{f}^\prime_{\mathrm{DMI}}(\varphi) - \nonumber \\ &\frac{\pi}{2}\left(H_y \cos \varphi-H_x \sin \varphi\right)\bigg)\bigg],  \\ \dot{\Delta}=&\frac{12 \gamma_0}{\mu_0 M_s \alpha \pi^2}\bigg[\frac{A}{\Delta}-\Delta\left(K_0+K \sin ^2 \varphi\right)+ \nonumber \\ &\mu_0 M_s \Delta \frac{\pi}{2}\left(H_x \cos \varphi+H_y \sin \varphi\right)\bigg]. \label{eq:DeltaDot}
\end{align}

Where we define
\begin{equation}
    H_K= \frac{2 K}{M_s \mu_0} ~,~ \tilde{f}^\prime_{DMI}(\phi) = \frac{1}{2 \Delta M_s \mu_0} \frac{\partial f_{DMI}(\phi)}{\partial \phi} 
\end{equation}

and $f_{DMI}(\phi)$ represents the trigonometric function with all the different DMI contributions (antisymmetric $D_a$, symmetric $D_s$, traceless $D_b$ and diagonal $D_t$)
\begin{equation}
    f_{DMI}(\phi) = (D_a - D_s)\cos \phi + (D_t - D_b) \sin \phi
\end{equation}
If we assume an up-down initial configuration ($Q = +1$) and an in plane (IP) field free stationary case (i.e $H_x = H_y = 0$), imposing the stationary conditions $\dot \phi = \dot \Delta = 0 $ yields the conditions \cite{Glathe2008,Fattouhi2022} for rigid motion of the DW magnetization
\begin{align}
    H_z &= \alpha\bigg( H_k ~ \frac{\sin 2 \phi}{2} - \frac{\pi}{2} \big( (H_{DMI,a} - H_{DMI,s})\sin \phi \nonumber \\ &+ ~ (H_{DMI,b} - H_{DMI,t})   \cos \phi\big) \bigg), \label{eq:walker_condition}
\end{align}
where $H_{DMI,i \in \{a,s,b,t\}} = D_i / 2 \Delta \mu_0 M_s $ is the effective field strength associated to the different DMI components. In order to make the notation more compact, we define 
\begin{equation}
    \kappa := \frac{K}{K_0} ~ , ~ \tilde{D}^\prime := \frac{\pi (D_a - D_s)}{\mu_0 H_K M_s \Delta_0} ~ , ~ \tilde{D}^{\prime\prime} :=  \frac{\pi (D_t - D_b)}{H_K \mu_0 M_s \Delta_0}  
\end{equation}
where $\Delta_0 = \sqrt{\frac{A}{K_0 + K \sin^2 \phi}} $ represents the equilibrium DW width that can be obtained by setting $\dot \Delta = 0$ in eq.\eqref{eq:DeltaDot}. These definitions allow us to rewrite \eqref{eq:walker_condition} in the form 
\begin{equation}
H_z  = \frac{\alpha H_k}{2} \left[ (\tilde{D}^{\prime\prime}  \cos \phi - \tilde{D}^{\prime} \sin \phi)\frac{\sqrt{1 + \kappa \sin^2 \phi}}{\kappa} + \sin 2\phi \right] \label{eq:walker_condition_2}.
\end{equation}
For fixed $\kappa , \tilde{D}^{\prime\prime} , \tilde{D}^{\prime} $, the Walker field is identified as the largest $H_z$ fulfilling eq.\eqref{eq:walker_condition_2} and is obtained by maximising the right hand side of eq.\eqref{eq:walker_condition_2} \cite{Li2022}, i.e. 
\begin{align}
    H_W &:= \frac{\alpha H_k}{2} \times \nonumber \\ &\max_{\phi \in [0,2 \pi) } \left[ (\tilde{D}^{\prime\prime}  \cos \phi - \tilde{D}^{\prime} \sin \phi)\frac{\sqrt{1 + \kappa \sin^2 \phi}}{\kappa} + \sin 2\phi \right] \label{eq:walker_Field}
\end{align}
The maximization of \eqref{eq:walker_Field} is not possible in closed analytical form, however one can treat the thin film limit, where the perpendicular magnetic anisotropy dominates over the shape anisotropy , i.e. $N_z \gg N_x ~,~ N_y$ implying the condition $\kappa \ll 1$ on eq.\eqref{eq:walker_Field}. The asymptotic solution in that case has the following form
\begin{align}
    H_W &\sim \begin{cases} \frac{\tilde{D}^{\prime\prime} ~ |\tilde{D}^{\prime\prime}| + |\tilde{D}^{\prime}|~ \tilde{D}^{\prime}}{\kappa \sqrt{(\tilde{D}^{\prime\prime})^2 + (\tilde{D}^{\prime})^2}}\text{ if } sign(\tilde{D}^{\prime\prime} \cdot \tilde{D}^{\prime}) = 1 \\ \frac{\tilde{D}^{\prime\prime} ~ |\tilde{D}^{\prime\prime}| - |\tilde{D}^{\prime}|~\tilde{D}^{\prime}}{\kappa \sqrt{(\tilde{D}^{\prime\prime})^2 + (\tilde{D}^{\prime})^2}} \text{ if } sign(\tilde{D}^{\prime\prime} \cdot \tilde{D}^{\prime}) = -1 \end{cases} \nonumber \\  &(\text{as } \kappa \rightarrow 0). \label{eq:Walker_analytic}
\end{align}
We validate the assumption of $\kappa \ll 1$ approximation in our case by pointing out how the demagnetizing factors in the case of a slab geometry can be calculated analytically \cite{Aharoni1998} and our geometry $L_x = 1 ~ \mu$m, $L_y = 160$ nm and $L_z = 0.6$ nm yields the following values for the demagnetizing factors
\begin{equation}
    N_x = 0.0013 ~ , ~ N_y = 0.0082 ~ , ~ N_z = 0.990 .
\end{equation}
We now proceed and discuss the obtained analytical result by comparing them with numerical simulations. By observing eq.\eqref{eq:Walker_analytic}, we first of all notice how in the limit of $\tilde{D}^{\prime} \rightarrow 0$ (i.e. a DMI tensor with only elements on the diagonal) or the limit $\tilde{D}^{\prime\prime} \rightarrow 0$ (i.e. a DMI tensor with only elements on the off-diagonal) the asymptotic behavior of eq.\eqref{eq:Walker_analytic} becomes 
\begin{align}
    H_W(\tilde{D}^{\prime} \rightarrow 0) &\sim \tilde{D}^{\prime\prime} / \kappa \label{eq:Walker_analytic_Db} \\
    H_W(\tilde{D}^{\prime\prime} \rightarrow 0) &\sim \tilde{D}^{\prime} / \kappa \label{eq:Walker_analytic_Dl} \\
\end{align}

\begin{figure*}[hbt]
    \noindent
    \includegraphics[width=\textwidth]{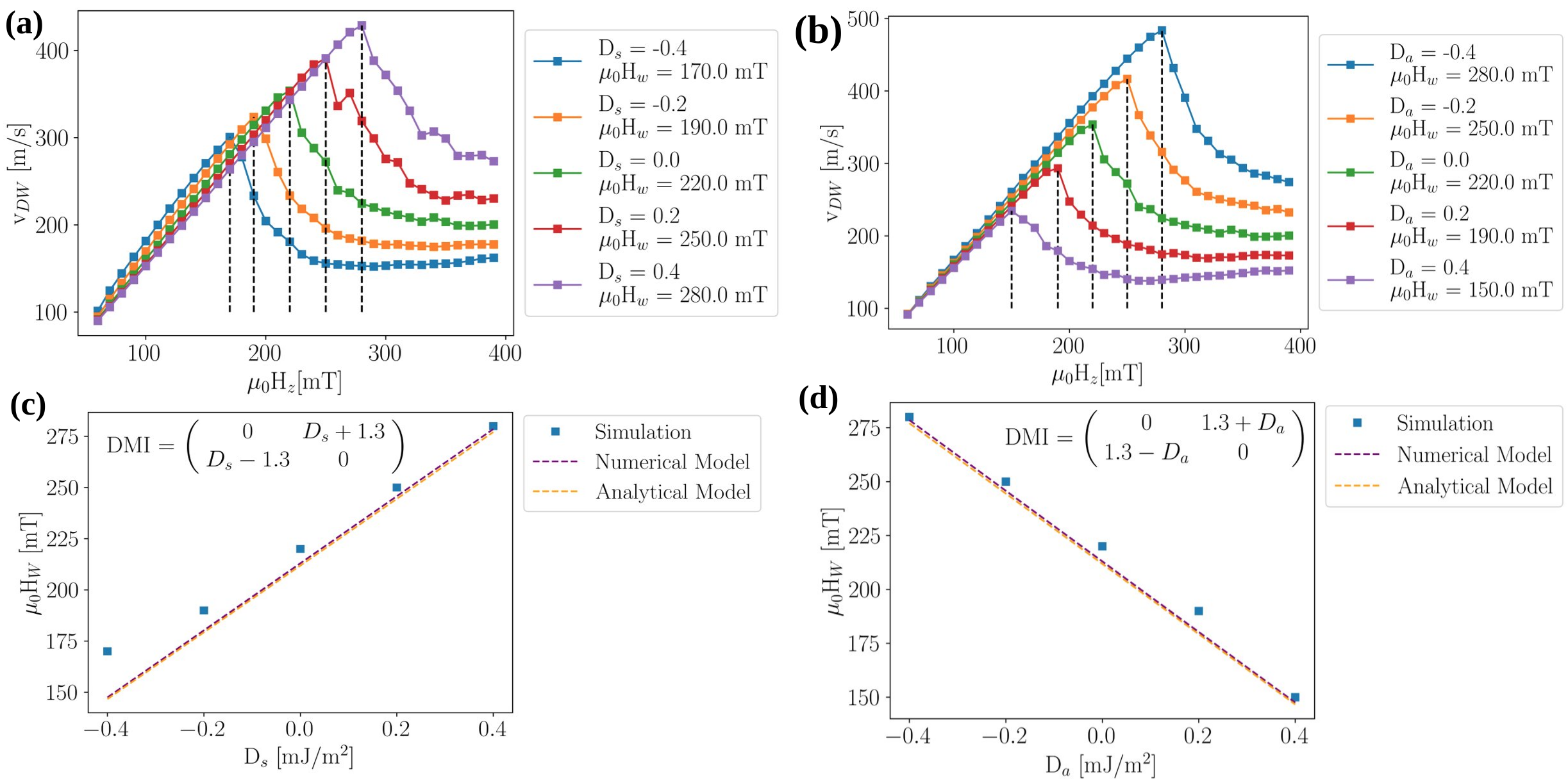}
    \caption{Comparison of Micromagnetic simulations \cite{Vansteenkiste2014}, numerical minimization  \eqref{eq:walker_condition} and analytical estimate (see eq.\eqref{eq:Walker_analytic}) for the DW Walker breakdown $H_W$. (a) DW velocity $v_{DW}$ as a function of an applied out-of-plane field $H_z$. The different curves show the velocity profile for different values of the symmetric component of a C$_{2v}$ symmetric DMI tensor (see inset of (c)). (b) DW velocity $v_{DW}$ as a function of an applied out-of-plane field $H_z$. The different curves show the velocity profile for different values of the anti symmetric component of a C$_{2v}$ symmetric DMI tensor (see inset of (d)). (c) Comparison of simulation results, numerical maximization of eq.\eqref{eq:walker_condition_2} and analytical estimate (see eq.\eqref{eq:Walker_analytic}) of the Walker breakdown as a function of the symmetric component of a C$_{2v}$ symmetric tensor. (d) Comparison of simulation results, numerical maximization of eq.\eqref{eq:walker_condition_2} and analytical estimate (see eq.\eqref{eq:Walker_analytic}) of the Walker breakdown as a function of the anti symmetric component of a C$_{2v}$ symmetric tensor.   }
    \label{fig:WB_Ds_Db}
\end{figure*}
\begin{figure*}[hbt]
    \noindent
    \includegraphics[width=\textwidth]{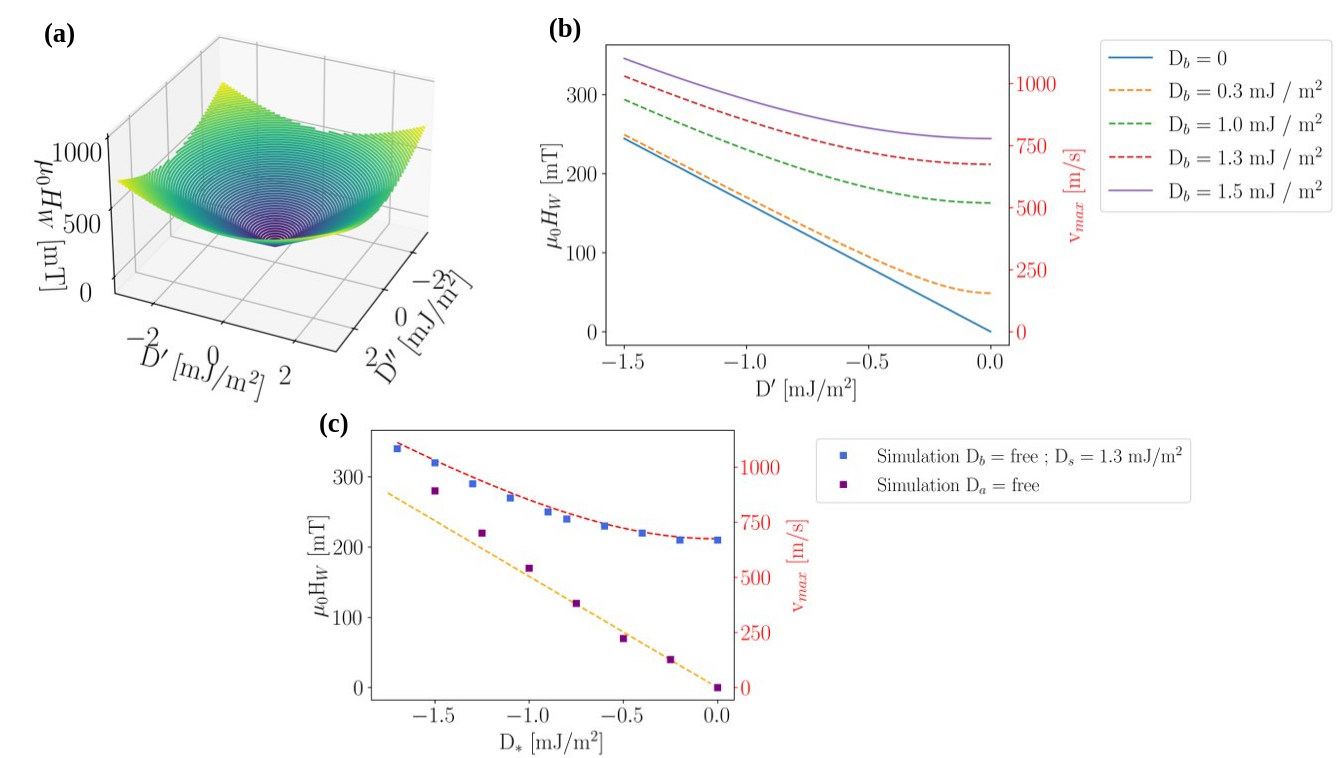}
    \caption{(a) 3D plot of eq.\eqref{eq:Walker_analytic} (b) Plot of the analytical formula for the WB field $H_W$ as a function of the off-diagonal $\tilde{D}^{\prime}$ component (see eq.\eqref{eq:Walker_analytic}). The different curves represent the behavior of the WB field for different diagonal DMI $D_b$ values. At $D_b = 0$ (i.e the blue curve) we recover the linear behavior. The peak velocities are reported on the second y-axis shown in red.  (c) Comparison of WB field calculated from a micromagnetic simulation with a DMI tensor compatible with $S_4$ symmetry  ($D_l = 1.3$ mJ/m$^2$ and free $D_b$) and the analytical estimate of eq.\eqref{eq:Walker_analytic}. The peak velocities are reported on the second y-axis shown in red. }
    \label{fig:WB_non_lin}
\end{figure*}

which shows a linear behavior compatible both with our numerical results (see Fig.\ref{fig:WB_Ds_Db}) and, in the  $H_W(\tilde{D}^{\prime} \rightarrow 0) \sim \tilde{D}^{\prime\prime} / \kappa $ case, with the results shown in \cite{Li2022}. We emphasize how these limiting cases show a linear dependence of the WB field only in the case of exclusive presence of diagonal or off-diagonal elements, but not both at the same time. By observing Fig.\ref{fig:WB_non_lin}-(a,b) we point out how the presence of both a diagonal and off-diagonal component of the DMI tensor results in a departure from the linear behavior also described by eq.\eqref{eq:Walker_analytic_Db} and eq.\eqref{eq:Walker_analytic_Dl} hinting at the fact that the components of the effective field counteracting precessional motion do not cooperate additively but in a non-linear way. Furthermore, we emphasize how this behavior of the WB field directly translates in the attainable peak DW velocity since, in the $\kappa \ll 1$ limit, 
\begin{align}
&v_{max} \sim \frac{\Delta_0 \gamma_0 \alpha}{1 + \alpha^2} H_W = \\ &\frac{\Delta_0 \gamma_0 \alpha}{1 + \alpha^2} \begin{cases} \frac{\tilde{D}^{\prime\prime} ~ |\tilde{D}^{\prime\prime}| + |\tilde{D}^{\prime}|~ \tilde{D}^{\prime}}{\kappa \sqrt{(\tilde{D}^{\prime\prime})^2 + (\tilde{D}^{\prime})^2}}\text{ if } sign(\tilde{D}^{\prime\prime} \cdot \tilde{D}^{\prime}) = 1 \\ \frac{\tilde{D}^{\prime\prime} ~ |\tilde{D}^{\prime\prime}| - |\tilde{D}^{\prime}|~\tilde{D}^{\prime}}{\kappa \sqrt{(\tilde{D}^{\prime\prime})^2 + (\tilde{D}^{\prime})^2}} \text{ if } sign(\tilde{D}^{\prime\prime} \cdot \tilde{D}^{\prime}) = -1 \end{cases} \nonumber \\ & (\text{as } \kappa \rightarrow 0). \label{eq:peak_vel}  
\end{align}
In Fig.\ref{fig:WB_non_lin}-(c) we report the peak velocities calculated with eq.\eqref{eq:peak_vel} and show how with $D_s = 1.5$ mJ/m$^2$ and $D_b = -1.5$ mJ/m$^2$ even peak velocities as high as $v_{max} \approx 1200 $ m/s are theoretically achievable. Furthermore, using experimentally measured \cite{Karube2021} parameters for the $S_4$ symmetric Schrebersite compound  Fe$_{1.9}$Ni$_{0.9}$Pd$_{0.2}$P ($A = 8 ~ \text{pJ/m} ~,~ K_u =31~ \text{kJ/m}^3 ~,~ M_s = 417 ~ \text{kA/m} $)
while keeping the nanowire dimensions unchanged, eq.\eqref{eq:peak_vel} predicts how peak velocities $v_{max} \approx 1700 $ m/s can be achieved even with much smaller DMI tensor components (i.e. $D_s = D_b = 0.2 $ mJ/m$^2$).  
\section{Conclusion}\label{Sec:8}
In this work we modified to the existing CCM \cite{Thiaville2002,Boulle2013,Nasseri2018} to include and study the effects of arbitrary DMI tensors on the statics and dynamics of domain walls in magnetic nanowires. We discuss how the effects of a DMI tensors can be described by inspecting how its symmetric traceless ($D_s, D_b$), antisymmetric ($D_a$) and diagonal ($D_t$) components act on the effective field inside the DW. We first show how DW canting is well described by the energy density of the $q - \phi - \chi$ model (see Fig.\ref{fig:simulations}) and discuss how the canting angle method is able to distinguish diagonal ($D_b$,$D_t$) DMI contributions and off-diagonal DMI contributions ($D_a,D_s$). We also observe how measuring the response of the canting angle $\chi$ to the simultaneous application of an IP field with both H$_x$ and H$_y$ components could potentially be a means to magneto-optically measure symmetric ($D_s$) and antisymmetric ($D_a$) contributions (see Fig.\ref{fig:Da_Ds_measures}) to DMI. Other IP field applications schemes could be studied to further enhance the resolution power of this technique. We then proceed and show how, in the presence of both $D_s$ and $D_b$ DMI components, DW tilting can be present even in the absence of IP fields. We derive a simple analytic formula for the canting angle $\chi$ as a function of $D_b$ valid in the $D_s/D_b \ll 1$ limit (see Fig.\ref{fig:Intrinsic_DMI}). We then study the effect of the different DMI tensor components on the the field driven dynamic properties of DWs in magnetic nanowires. We discover that the effects of the interplay of the Néel- and Bloch-Stabilizing DMI components on the magnitude of the WB field is not trivial and determines a departure from the simple linear dependency (Fig.\ref{fig:WB_non_lin}-(a)) in the case of pure interface- \cite{Thiaville2012} of bulk-DMI \cite{Li2022}. We then derive an analytic formula describing the dependency the of the WB field on the different DMI tensor components (eq.\eqref{eq:Walker_analytic}) in the thin film limit, comparing its predictions with micromagnetic simulations (Fig.\ref{fig:WB_non_lin}-(b)). The very high theoretically achievable DW velocities in the order km/s (Fig.\ref{fig:WB_non_lin}-(c)) is confirmed by simulation and could open the way to a new wave of experimental investigations in low-symmetrty magnetic thin films. These results indeed hint at the fact that materials displaying these more exotic forms of DMI combining both Bloch and Néel stabilising effective fields, could be interesting candidates for novel DW motion based technology concepts.

\section{Acknowledgement}
This project has received funding from the European Union’s Horizon 2020 research and innovation programme under the Marie Skłodowska-Curie grant agreement No. 860060 “Magnetism and the effect of Electric Field” (MagnEFi).
\renewcommand\refname{References}
%
\newpage

\end{document}